\documentclass[aip,apl,reprint,superscriptaddress,amsmath,amssymb]{revtex4-1}
\usepackage{epsfig,psfrag}
\usepackage{dcolumn}
\usepackage{bm}
\usepackage{graphicx}
\usepackage{color}

\begin{document}
\title{Relevance of intra- and inter-subband scattering on the absorption in heterostructures}

\author{C. Ndebeka-Bandou}
\author{F.Carosella}
\author{R. Ferreira}
\affiliation{Laboratoire Pierre Aigrain, Ecole Normale Sup\'erieure, CNRS (UMR 8551), Universit\'e P. et M.Curie,
Universit\'e D. Diderot, 24 rue Lhomond F-75005 Paris, France}

\author{A. Wacker}
\affiliation{Mathematical Physics, Lund University, Box 118, S-22100 Lund, Sweden} 

\author{G. Bastard}
\affiliation{Laboratoire Pierre Aigrain, Ecole Normale Sup\'erieure, CNRS (UMR 8551), Universit\'e P. et M.Curie,
Universit\'e D. Diderot, 24 rue Lhomond F-75005 Paris, France} 
\affiliation{Technical University Vienna, Photonics Institute, Gusshausstrasse 27, A-1040 Vienna, Austria}

\begin{abstract} 
We analyze the absorption lineshape for  inter-subband
transitions in disordered quasi two-dimensional heterostructures by an exact
calculation. The intra-subband scatterings control the central  peak while
the tails of the absorption line are dominated by the inter-subband scattering
terms. Our numerical study  quantitatively assesses the magnitude of the free
carrier absorption. The accuracy of different models currently used for
gain/absorption is discussed.
\end{abstract}

\pacs{73.21.Ac,78.67.Pt}

\maketitle

The demand to produce reliable terahertz and infrared devices has
triggered extensive researches  on intersubband transitions in semiconductor
heterostructures \cite{ref2}.~A highlight is the realization and  continuous
improvement of  Quantum Cascade Lasers (QCL) \cite{ref1,ref4,ref5,ref6}
that emit in this frequency range.  The search for improved performances of
QCLs includes a better design of the layer sequence to enhance the population
inversion between the two subbands involved in the lasing transition, the
decrease of the non radiative paths, but also the control of the photon
re-absorption by the free carriers that are present in the structure
\cite{ref7,ref8,ref9,ref10,ref11,ref12,ref13}.  The free carrier absorption
(FCA), well known from bulk structures, is transferred to a variety of
intra-subband and inter-subband oblique (in two-dimensional $\vec{k}$ space) transitions
in heterostructures\cite{ref15,ref14}. Tailoring  these (often parasitic)
processes requires a thorough understanding of  the nature of the optical
transitions in imperfect heterostructures.\\

For light polarized along the growth ($z$) axis of an ideal semiconductor
heterostructure, the energy dependence of the absorption coefficient
between  subbands 1 and 2 is a Dirac delta function line  centered at
$\hbar\omega_0=E_2-E_1$ \cite{ref16}. This ``atomic-like'' profile stems from
the fact that the electromagnetic wave coupling  requires a non-vanishing
$z$-dipole $\langle1|\hat{p}_z|2\rangle$, but does not affect the in-plane
motion of the electrons.  These motions are identical in the two subbands
(i.e. plane waves, owing to the in-plane translation invariance) and their
corresponding energies form identical dispersions (if we assume the same
parabolic mass $m^*$). Static disorder affects the in-plane motion by
introducing intra-subband ($V_\mathrm{def}^\mathrm{intra}$) as well as
inter-subband ($V_\mathrm{def}^\mathrm{inter}$) couplings. Bound states
usually appear below the subband edges.  As a consequence, the strict
selection rules leading to the Dirac lineshape are no longer expected to apply
to a disordered  structure. This is frequently taken into account by
some effective broadening. Here we treat this feature from a fully microscopic
point of view by considering the exact eigenstates of the disordered quantum
well potential. Our calculations allow us to gain a complete insight into
the  underlying physical processes leading to the absorption in
heterostructures,  not only near resonance, but also far  above or
far below the resonance (theses regions have not deserved enough attention in
the literature, even though of great technological interest). We will show
that the  regions far from resonance can nicely be understood by resorting to
defect-induced  scattering processes, and highlight the respective roles
of  $V_\mathrm{def}^{\rm intra}$ and $V_\mathrm{def}^{\rm inter}$. Finally,
the complete calculation will allow us to determine the accuracy of different
models currently  employed to evaluate the gain/absorption of such
structures. 

In the presence of disorder, the heterostructure eigenstates 
can be written as $\Psi_\nu(\vec{\rho},z)=\sum_{n}F_{\nu,n}(\vec{\rho})\chi_n(z)$, with:
\begin{multline}
 \frac{-\hbar^2}{2m^*}\nabla^2F_{\nu,n}(\vec{\rho})+\sum_{n'}F_{\nu,n'}(\vec{\rho})
V_\mathrm{def}^{n,n'}(\vec{\rho})\\
=\left( \varepsilon_\nu-E_n\right)F_{\nu,n}(\vec{\rho})
\label{eq2}
\end{multline} where $\vec{\rho}$=$(x,y)$, $F_{\nu,n}(\vec{\rho})$ is the
in-plane envelope function of the $n^{th}$ subband and
$V_\mathrm{def}^{n,n'}(\vec{\rho})=\langle \chi_n(z)
|V_\mathrm{def}(\vec{\rho},z) | \chi_{n'}(z) \rangle$. $\chi_n(z)$ is the
heterostructure localized part and $E_n$ the corresponding bound state. In 
the absence of defects  $\nu\rightarrow(n,\vec{k})$,
$F_{\nu}(\vec{\rho})\rightarrow \exp(i\vec{k}\cdot\vec{\rho})/\sqrt{S}$ and
$\varepsilon_\nu\rightarrow E_n+\hbar^2k^2/(2m^*)$. In the following we shall
present results obtained after a numerical diagonalization of  Eq.~(\ref{eq2})
within a truncated basis with the two lowest subband states ($n$=1,2). In
practice, we expand $F_{\nu,n=1,2}(\vec{\rho})$ in a plane wave basis 
that fulfill periodic (Born-von Karman) conditions in a 200$\times$200 nm$^{2}$ box. The matrix
element of the dipole coupling to light are proportional to
$(\hat{p}_z)_{(\nu,\mu)}=\langle \Psi_\nu(\vec{r})|\hat{p}_z |
\Psi_\mu(\vec{r})\rangle$.  For defect-free structures one has the selection
rules:
$(\hat{p}_z)_{(n,\vec{k}),(n',\vec{k'})}=(1-\delta_{n,n'})\delta_{\vec{k},\vec{k'}}
\langle\chi_n|\hat{p}_z|\chi_{n'}\rangle$, reflecting the in-plane translation
invariance and the non-vanishing $z$-dipole.  In the following, we will be
interested in the effect of disorder on these matrix elements.

In QCLs the interface defects are important  elastic scatterers. Most features can be
directly transferred to other mechanisms such as impurity \cite{ref24} and alloy
scattering. Interface defects arise from the non ideality of the
well/barrier interface at $z$=$z_0$ between two consecutive layers of the
heterostructure: they can be a protrusion of the barrier material in the well
(repulsive defects) or vice-versa  (attractive defects). For nearly ideal
interfaces, the interface defects are only one monolayer thick.  The
scattering potential created by one disordered interface has the form
\begin{equation} 
V_\mathrm{def}(\vec{r})=f(z)u_\mathrm{def}(\vec{\rho})
\label{eq3}
\end{equation}
showing a separation on the dependence on the $z$ and $\vec{\rho}$
variables. $u_\mathrm{def}(\vec{\rho})$ is the superposition of localized
functions centered at the randomly placed interface defects on the $(x,y)$
plane \cite{ref14}. For a barrier/well  interface at $z$=$z_0$ and for a repulsive defect
$f^\mathrm{rep}(z)=V_b\Theta(z-z_0)\Theta(z_0+h-z)$ while for an  attractive defect
$f^\mathrm{att}(z)=-V_b\Theta(z_0-z)\Theta(z-z_0+h)$, where $V_b$ is the barrier
height, $h$ is the  thickness of one monolayer, $\Theta(z)$ is the Heaviside
function. 

\begin{figure}
 \includegraphics[scale=0.2]{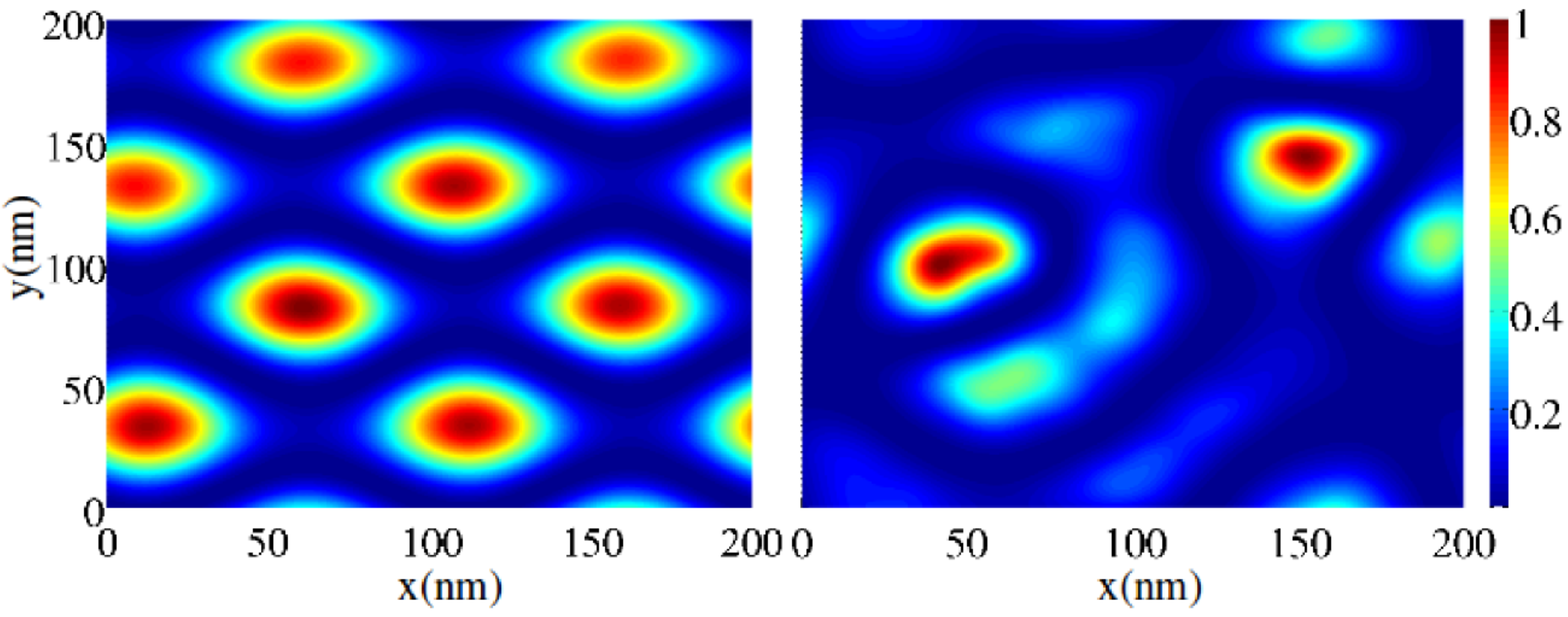}
 \caption{Color plot of the normalized in-plane probability distribution for two states with energies equal to $E_1$+2 meV (left panel) and $E_2$+2 meV (right panel). The quasi $E_1$ state is extended in the $(x,y)$ plane while the quasi $E_2$ state is fairly localized by the interface defects.}
\label{fig0}
\end{figure}

\begin{figure}
 \includegraphics[scale=0.27]{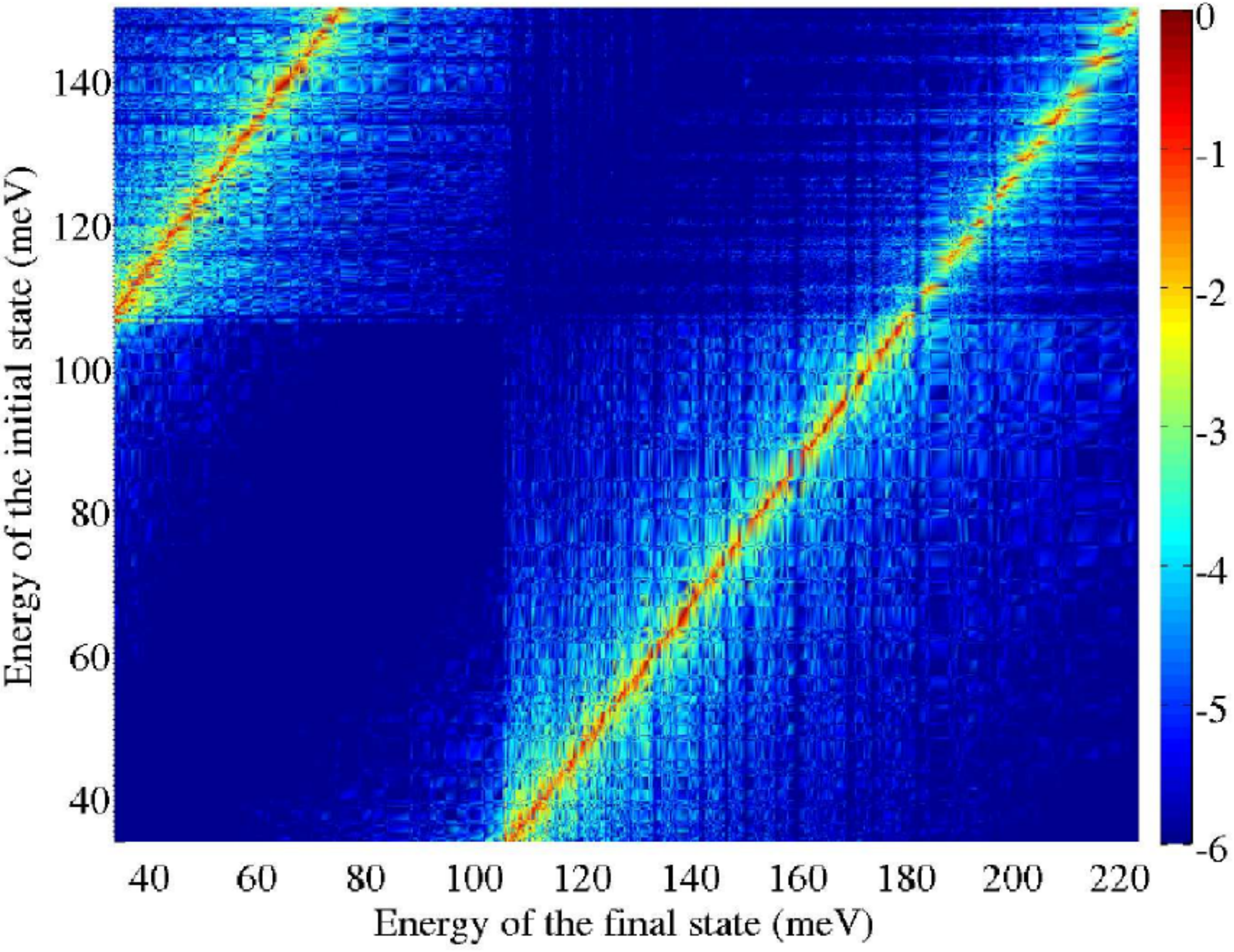}
 \caption{Color plot of the decimal logarithm of the normalized squared modulus of the optical matrix
elements for the ensemble of transitions $\varepsilon_\nu$-$\varepsilon_\mu$ in a disordered 9/2/3 nm
GaAs/Ga$_{0.75}$Al$_{0.25}$As double quantum well. The disorder is due to interface defects
randomly distributed on the two inner interfaces of the structure.}
\label{fig1}
\end{figure}

 Here
we consider a 9/2/3 nm GaAs/Ga$_{0.75}$Al$_{0.25}$As double quantum
well (DQW) structure with Gaussian interface defects \cite{ref14}
placed on the two inner interfaces,  with a characteristic in-plane
extension of 3.6 nm and a fractional coverage of the surface
$N_\mathrm{def}\pi \sigma^2/S=30\%$.  Fig.~\ref{fig0} shows the
in-plane probability distribution for two states with energies
respectively equal to $E_1+$2 meV and $E_2$+2 meV. The extra energies
+2 meV are larger (respectively smaller) than the typical effective
in-plane  potential depths in the $E_1$ and $E_2$ subbands (0.4 meV
and 6 meV respectively). Hence, these two in-plane electronic
distributions look very different: the quasi $E_1$ state is extended
in the $(x,y)$ plane and approximately given by the wave-function
  $\Psi(x,y)\propto \cos(k_0x+\varphi_x)\cos(k_0y+\varphi_y)$, where
  $\varphi_{x/y}$ are phases and $k_0=4\pi/200$ nm. Its kinetic energy
  $\hbar^2k_0^2/2m^*\approx 2$ meV is significantly larger than the
characteristic potential depth. In contrast, the ``mostly $E_2$
state'' is fairly localized by the interface defects. In
Fig.~\ref{fig1} we show the matrix $|\langle
\Psi_\nu(\vec{r})|\hat{p}_z | \Psi_\mu(\vec{r})\rangle|^2$ for our
calculated eigenstates  of the disordered heterostructure.  The figure
clearly displays two blurred straight regions around
$|\varepsilon_\mu-\varepsilon_\nu|\approx\hbar\omega_0=73.8$ meV,
corresponding to the subband spacing  in this sample. If there where
no disorder, there would be no blurring since (see above) a single
final state would match any given initial state. The fact that the
matrix element is almost zero if the energy difference between the
true states differs strongly from the intersubband spacing corresponds
well with the conventional broadening picture.

In this context it has to be noted that the blurring is strongly reduced, if
the wave functions $\chi_n(z)$ for the subbands $n=1$ and $n=2$ are of the
same magnitude at each interface. Neglecting $V^\mathrm{inter}_\mathrm{def}$
this provides identical (or very similar) in-plane wave functions for both
subbands, which provide strong selection rules for the $p_z$-matrix elements.
Similar selection rules appear in strong magnetic fields \cite{ref23}.  In our
case, the wave functions differ essentially at the interfaces, as the
excited state penetrates deeper into the barrier, so that this effect is
only of very minor relevance.

\begin{figure}
 \includegraphics[scale=0.27]{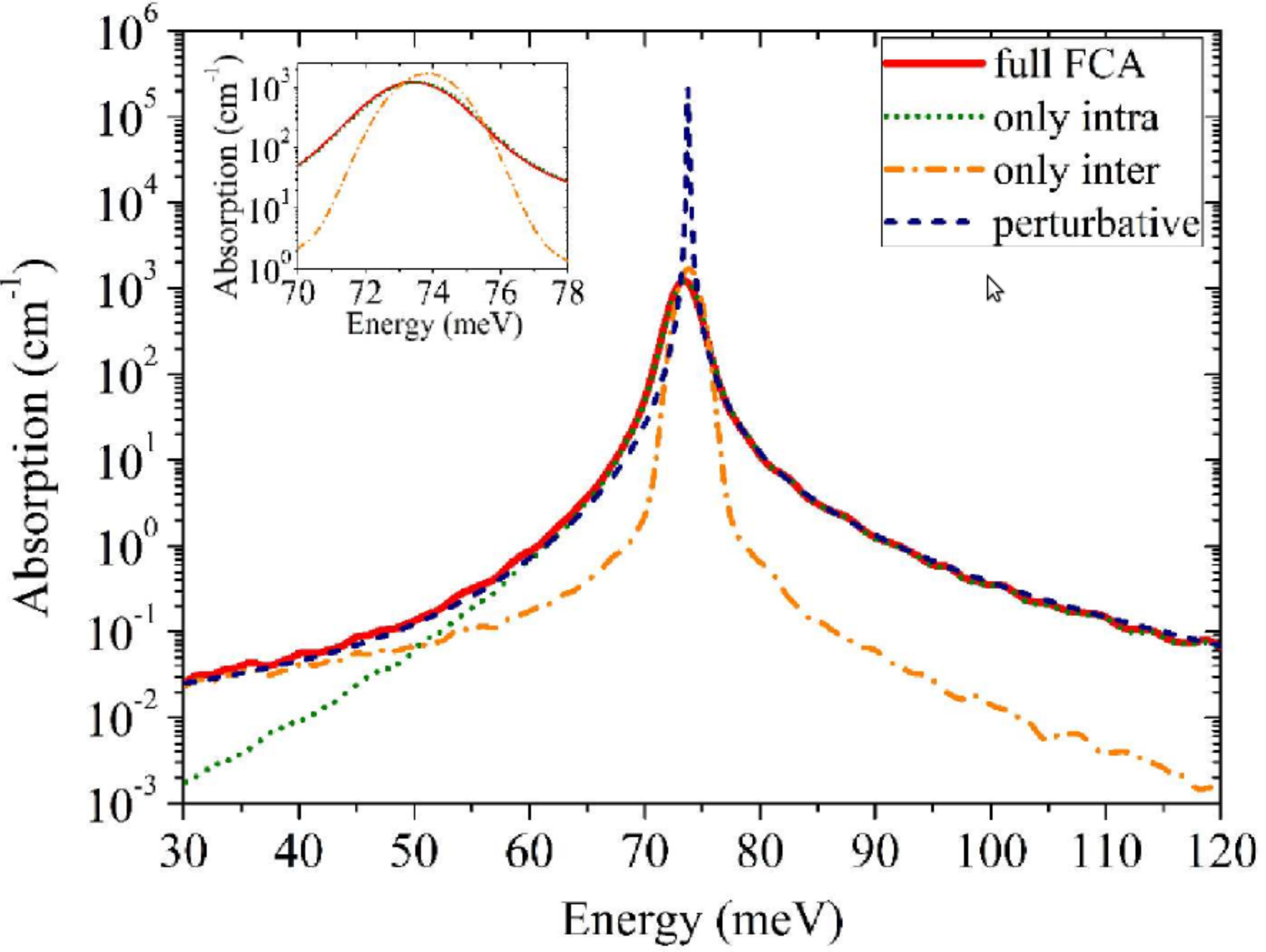}
 \caption{Absorption spectrum for the $E_1$-$E_2$ transition calculated by: fully
numerical diagonalization (red full line); taking into account  either only $V_\mathrm{def}^\mathrm{intra}$ (green dotted line) 
or only $V_\mathrm{def}^\mathrm{inter}$ (orange dashed-dotted line); 
and by expanding the electron wavefunction to the first order in both $V_\mathrm{def}^\mathrm{intra}$ and $V_\mathrm{def}^\mathrm{inter}$ (see Ref.~\onlinecite{ref14}) 
(blue dashed line). Inset: Zoom of the absorption spectrum in (a) around
$\hbar\omega_0$. $T$=$100$ K.}
\label{fig2}
\end{figure}

Assuming an ideal overlap between the photon modes with the angular frequency $\omega$ and the DQW structure, the absorption coefficient (in SI units) is given by :
\begin{multline}
 \alpha(\omega)=\frac{2\pi e^2}{m^{*2}\omega\varepsilon_0cnL_zS}\sum_{\nu,\mu}\left(f_\nu-f_\mu\right)
 \left|\langle \Psi_ \nu|p_z|\Psi_\mu \rangle  \right|^2 \\ \times \delta\left(\varepsilon_\mu - \varepsilon_\nu - \hbar\omega \right)
\end{multline}
where the thickness of one period $L_z$ is equal to 19.6 nm, $n$ the refractive index and $f_\nu$ the occupation of the state $|\nu\rangle$ with energy $\varepsilon_\nu$.
We show in Fig.~\ref{fig2} (red full line) the absorption coefficient of
the DQW structure based on the matrix elements of Fig.~\ref{fig1} (in all
absorption calculations, the delta of energy conservation was replaced by a
Gaussian with a FWHM=1.88 meV).  In this absorption spectrum, the
contributions of $V_\mathrm{def}^\mathrm{intra}$ and
$V_\mathrm{def}^\mathrm{inter}$ are taken at all orders and are fully
admixed. In order to highlight the physics underlying the assisted light
absorption and disentangle the various contributions to its spectrum, we
discuss in the following  the results obtained within different approximations
and/or models. We also show in Fig. \ref{fig2} the calculated absorptions when
only $V_\mathrm{def}^\mathrm{intra}$ (green dotted line) or only
$V_\mathrm{def}^\mathrm{inter}$ (orange dashed-dotted line) are retained in
the scattering matrix. We see that the low energy (the high energy) FCA is
dominated by the $V_\mathrm{def}^\mathrm{inter}$
($V_\mathrm{def}^\mathrm{intra}$) couplings. Indeed, the initial and final
states involved in the non-resonant absorption displays a dominant $E_1$ or
$E_2$ subband  character. Thus, since the dipole coupling only triggers
inter-subband events and starting from an initial state in $E_1$, an
additional inter-subband scattering is needed for the low energy intra-$E_1$
oblique transition.  Conversely, the high-energy absorption tail involves
essentially $E_1$-to-$E_2$ processes and thus rely mostly on
$V_\mathrm{def}^\mathrm{intra}$.  It is worth pointing out that the
near-resonant absorption (see Fig.~\ref{fig2}) is accurately accounted for by the
only-intra approximation,  whereas the only-inter one fails in predicting both
the intensity and the absorption profile.  Finally, we also show in
Fig.\ref{fig2} the perturbative estimates of the FCA obtained by expanding the
electron wavefunction to the first order in $V_\mathrm{def}$ \cite{ref14}
(dashed blue line). We see that the perturbative estimate provides an
excellent rendering of the shape and strength of the actual absorption when it
is justified, i. e. not too close to the resonance.

\begin{figure}
 \includegraphics[scale=0.27]{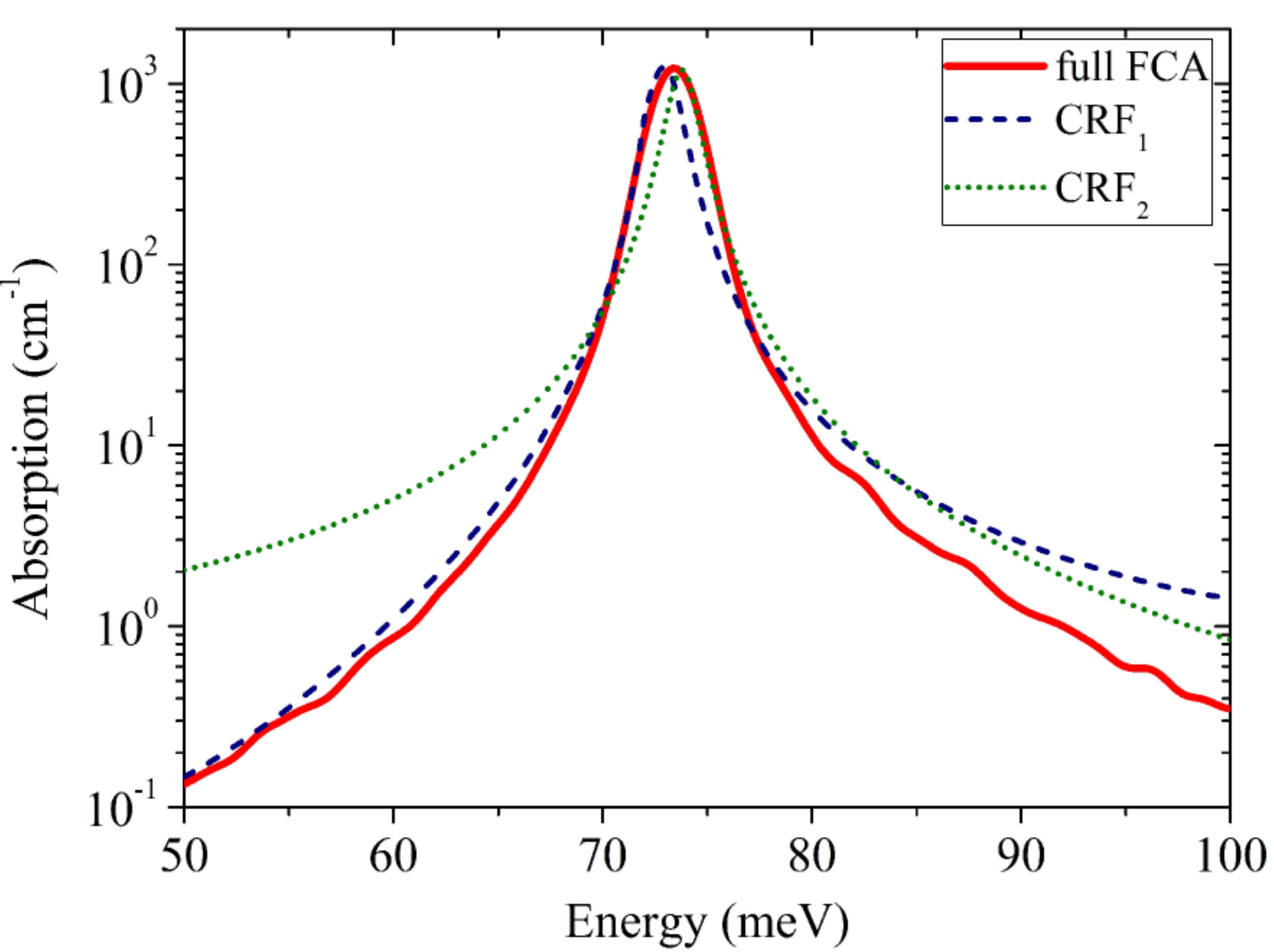}
 \caption{Absorption spectrum for the $E_1$-$E_2$ transition calculated by three different models:
exact diagonalization (red full line), Keldysh Green's function formalism, CRF$_1$ (blue dashed line), Unuma’s
model, CRF$_2$ (green dotted line). $T$=$100$ K.}
\label{fig3}
\end{figure}

An important issue is to understand how the results of the exact
diagonalization compare to correlation function approaches, essentially based
on  Green's function methods. To this end, we show in Fig.~\ref{fig3} the
absorption coefficient for the $E_1-E_2$ transition of the DQW structure
described above, calculated with two different correlation function (CRF)
approaches, and we compare it with the one calculated by exact
diagonalization.  In CRF$_1$ the Keldysh Green's function formalism was
applied using the implementation of Refs.~\onlinecite{ref19,ref20}. 
In order to make a quantitatively meaningful comparison among the various
calculations, we numerically extracted   the correlation length and the
average defect depth from the randomly generated interface defects used in the
exact calculation without the use of any fitting parameters. We see that
CRF$_1$ gives a good description of the low $\omega$ behavior, as the
inclusion of the nondiagonal self-energies  \cite{ref19} fully covers the
inter-subband terms addressed above. At large energy, CRF$_1$ overestimates
the absorption; a feature that is related to the use of $k$-independent
scattering matrix elements in order to simplify this numerics. This assumption
of an effective delta- potential scattering potential overestimates the
scattering for large wave-vector transfer, which is crucial for the tails. 
In CRF$_2$, we use
\textit{Unuma et al.} \cite{ref17} formalism for the intersubband absorption
coefficient that follows Ando's approach \cite{ref18}. To draw CRF$_2$ we
converted \textit{Unuma et al.}'s real part of the conductivity into an
absorption coefficient following \cite{ref14} and we used the interface
disorder potential described above. It is clear that CRF$_2$ poorly describes
the FCA far from resonance. This is most probably  due to the fact that,
contrarily to CRF$_1$, the off-diagonal components of the Green's function are
neglected, whereas the diagonal ones are evaluated within the self-consistent
Born approximation. For the peak, the CRF$_1$ and CRF$_2$ provide similar results
with a full width at half maximum of 1.8 meV and 1.6 meV, respectively,
without using  any adjustable parameters. This is also in good agreement with 
the value of 2.5 meV for the exact calculation (see Fig.~\ref{fig3}). Note that the actual width of the FCA model results from a convolution between the intrinsic broadening
 effects (caused by scattering in an infinite sample) and the numerical broadening that we have to input in the numeric to 
 account for the finite size of the sample in the simulation. A quantitative analysis of inter-subband linewidth should
 include additional scattering mechanisms (impurities, phonons) and is beyond the scope of this paper. 

In conclusion, we have computed numerically the lineshape of inter-subband
transitions in heterostructures with interface disorder. We have found
that only few final states are optically connected to a given initial state
despite disorder, which reflects the common picture of a broadened
transition.  While intrasubband scattering dominates at the resonance peak
and at higher photon energies, intersubband scattering dominates the low
energy tail of the absorption spectrum.  We have compared the outputs of
several models to compute absorption to the fully numerical results, in
particular their low $\omega$ behaviors.  The perturbative estimate for FCA
\cite{ref14} works very well for the absorption tails.  We find also that
\textit{Unuma et al.}'s model \cite{ref17} gives a poor description of the
non-resonant FCA tails, while the Keldysh Green's function formalism
\cite{ref19}  with full nondiagonal self-energies  fits the numerical results
nicely. A clear picture of the inter-subband and intra-subband transitions now
emerges from this study and emphasizes the need for a proper account of the
perturbation of the carrier wavefunctions by defects and not only of their
energies.
\begin{acknowledgements}
We thank E. Dupont, K. Unterrainer and G. Strasser for valuable
discussions. A. Wacker thanks the Swedish Research Council (VR) 
for financial support.
\end{acknowledgements}

\bibliographystyle{apsrev4-1}

%

\end{document}